\documentclass[11pt]{article}
\usepackage{mycite}

\def\.{{\cdot}}
\def\gtapprox{\,\lower.6ex\hbox{$\buildrel >\over \sim$} \, }
\def\ltapprox{\,\lower.6ex\hbox{$\buildrel <\over \sim$} \, }

\def\arcs{\ifmmode {'' }\else $'' $\fi}     
\def\arcm{\ifmmode {' }\else $' $\fi}     
\def\deg{\ifmmode^\circ\else$^\circ$\fi}    
\def\ttimes{{\scriptstyle \times}}
\def\fr7{7$ \hskip -0.9ex \vrule height0.8ex width0.8ex depth-0.73ex
                                                                \hskip0.1ex$}

\def\hMpc{~$h^{-1}$Mpc}

\newcommand\joref[5]{#1, #5, {#2, }{#3, } #4}

\def\apj{Ap.J.}                 

\def\aanda{A.\&A.}            

\title{M\'ethodes aux grandes \'echelles pour mesurer la topologie 
(globale) de l'Univers}

\author{Boudewijn Roukema\\
Observatoire de Strasbourg, 
11, rue de l'Universit\'e, Strasbourg 67000, 
France\\roukema@iap.fr}

\date{}

\begin{document}
\maketitle

Si l'Univers suit en moyenne la m\'etrique 
de Friedmann-\-Lema\^{\i}tre-\-Robertson-\-Walker, sa partie spatiale
(l'hypersurface extrapol\'ee \`a l'\'epoque actuelle \`a $t=t_0$ 
en coordonn\'ees commobiles) 
peut avoir soit 
une courbure non-nulle, soit une topologie non-triviale, 
soit toutes les deux (\cite{deSitt17}~1917; \cite{Lemait58}~1958).
Dans la suite, les mots $\ll$ courbure $\gg$ et $\ll$ topologie $\gg$ 
s'appliquent uniquement \`a cette hypersurface spatiale de l'Univers, 
en admettant que le Postulat de Weyl est satisfait par les 
observations astronomiques.

Les propri\'et\'es des galaxies, des quasars, du fond diffus cosmologique 
et d'autres objets vus  aux distances cosmologiques par des t\'elescopes 
au sol et spatiaux ne permettent pas encore 
de d\'eterminer sans \'equivoque le signe de 
la courbure de l'Univers. Elle est soit n\'egative, soit nulle, 
soit positive avec un rayon de courbure des dizaines d'ordres de magnitude
plus grand que l'horizon d\'efini par l'\^age fini de l'Univers.

L'hypoth\`ese de la plupart des cosmologistes observationnels 
selon laquelle 
la topologie de l'Univers est triviale implique donc que le volume 
(spatial) de l'Univers est soit infini (en cas de courbure non-positive), 
soit fini, mais est $10^{3N}$ fois le volume observable, o\`u $N\gg10.$ 

Selon cette hypoth\`ese, plus de 99,999\dots\% du volume de l'Univers est
impossible \`a observer.

Par contre, une approche empirique de la topologie de l'Univers pourrait
d\'emontrer que l'Univers est fini et observable dans sa grande partie. 

Pour une revue d\'etaill\'ee sur les techniques 
traditionnelles de d\'etection,
ou de mesure, de la topologie cosmologique, voir \cite{LaLu95}~1995 (ou
\cite{Lum98}~1998).

Quelques-unes des techniques les plus r\'ecentes sont pr\'esent\'ees ici.
Celles-ci consistent en des
m\'ethodes utilisant l'information 
\`a trois dimensions (\cite{LLL96}~1996; \cite{Rouk96}~1996;
\cite{RE97}~1997) ou \`a deux dimensions (sur le fond diffus cosmologique~: 
\cite{Corn96b}~1996b).

\section{Techniques tridimensionnelles : images topologiques}

Le principe de base de d\'etection de la topologie est que l'on verrait
les m\^emes objets (ou points d'espace) dans plusieurs directions 
angulaires du ciel et aux distances apparentes 
(d\'ecalages spectraux vers le rouge, 
$z$) tr\`es diff\'erentes. Mais puisque nous voyons des objets dans le 
pass\'e, il faut prendre en compte 
que les {\em images topologiques} distantes
seraient celles des objets vus dans leur jeunesse, voire avant qu'ils
soient n\'es. L'identification de deux images topologiques d'un seul 
objet pourrait donc \^etre tr\`es difficile.
Ces probl\`emes, qui sont li\'es \`a la formation et \`a l'\'evolution des
objets, sont les m\^emes que pour l'\'etude des autres param\`etres de la 
g\'eom\'etrie cosmologique ($\Omega_0, H_0, \Lambda$).

Pour comprendre les m\'ethodes de d\'etection, il faut avoir une
certaine intuition de ce qu'est un univers de topologie non-triviale.

L'on pourrait normalement imaginer un tel univers en commen\c{c}ant avec
un {\em poly\`edre fondamental} (domaine de Dirichlet) dont on identifie
des paires de faces d'une certaine fa\c{c}on. 
Les poly\`edres et les fa\c{c}ons
d'identifier des faces ne sont pas al\'eatoires, ils sont d\'etermin\'es
math\'ematiquement.  Il faut en m\^eme temps imaginer plusieurs copies du
poly\`edre qui
 $\ll$ chiffonnent $\gg$ le {\em rev\^etement universel}, c'est-\`a-dire, qui
recouvrent $R^3$ ou $H^3,$ pour une courbure nulle ou n\'egative
respectivement. Puisque l'\^age de l'Univers est fini, il suffit
d'imaginer assez de copies du poly\`edre pour que 
la sph\`ere observable (qui est finie) soit recouverte.

Pour tasser un espace de courbure n\'egative dans notre
intuition euclidienne, il faut se rappeler que plus une copie du
poly\`edre est loin du centre de la sph\`ere, plus elle semble
petite.

Dans toutes les m\'ethodes qui ne supposent pas a priori une topologie
quelconque, il faut donc r\'efl\'echir dans ce rev\^etement universel, qui
est (dans le cas de topologie non-triviale) uniquement un espace 
apparent dans lequel nous voyons plusieurs fois le (seul) 
Univers physique.

Pour l'espace euclidien sans constante cosmologique 
(not\'e $\Omega_0=1, \lambda_0=0$), le rayon 
de la sph\`ere observable (l'horizon) est $R_H = $6000\hMpc, o\`u 
$h= H_0/100$kms$^{-1}$Mpc$^{-1}$ est la constante de Hubble, de
valeur d'environ $0\.5-0\.8$. (Uniquement les coordonn\'ees commobiles
sont utilis\'ees ici, et les distances sont calcul\'ees pour $\Omega_0=1$ 
sauf indication contraire.)

Le rayon de la sph\`ere la plus petite dans laquelle on peut inscrire le
poly\`edre est $r_{>}$ et le rayon de la sph\`ere la plus grande que 
l'on peut inscrire \`a l'int\'erieur du poly\`edre est $r_{<}.$ 
Par d\'efinition donc, $r_> > r_<,$ et il est plus facile de contraindre
$r_>$ que $r_<,$ en particulier puisque la poussi\`ere dans 
le plan de notre propre galaxie
cache un dixi\`eme ou plus du ciel. (Cette fraction 
varie en fonction de la longueur d'onde \'electromagn\'etique 
utilis\'ee et du degr\'e d'incompl\'etude accept\'e pour un sondage.)

Les anciennes contraintes de limite inf\'erieure \`a $2r_>$ sont de l'ordre
de 60{\hMpc} \`a 150{\hMpc}.  Par exemple, \cite{Gott80}~1980 a 
mis les contraintes sur d'\'eventuelles images topologiques de 
l'amas de galaxies Coma, et l'existence 
de structures $\ll$ \`a grandes \'echelles $\gg$ 
(\cite{deLapp86}~1986; \cite{GH89}~1989;
\cite{daCosta93}~1993; \cite{Deng96}~1996; \cite{Einasto97}~1997) 
nous donnent des contraintes fortes \`a ces \'echelles 
que $2r_{>} \gtapprox R_H/100.$ 
En plus, il serait difficile de mettre une 
valeur de $2r_<$ aussi petite que $2r_< \ltapprox R_H/100$  en accord 
avec les observations.

Pour contraindre ou d\'etecter la topologie aux \'echelles de plus 
de $R_H/10,$
le probl\`eme de l'\'evolution et de la formation des objets astrophysiques
(galaxies, quasars, amas de galaxies) devient plus fort, et les 
catalogues d'objets observ\'es restent souvent limit\'es aux petites
distances ou aux petits angles solides. Les prochaines d\'ecennies
vont voir rapidement augmenter le nombre de catalogues plus 
$\ll$ profonds $\gg$\footnote{Attention : 
selon l'usage astronomique, 
$\ll$ profond $\gg$ veut dire que les objets vus sont 
faibles en magnitude apparente, ce qui 
n'implique pas forc\'ement que ces objets sont \`a une grande distance, 
puisque ils peuvent \^etre proches et de faible luminosit\'e intrins\`eque.}
sur une grande fraction du ciel.

Puisque la recherche d'images topologiques 
individuelles dans des catalogues aux \'echelles de plus de 10\% de
l'horizon souffre des effets de l'\'evolution des objets, des techniques
statistiques sont les plus simples. Les techniques de $\ll$ cristallographie 
cosmique $\gg$ (\S\ref{s-cristal}) et de la $\ll$ recherche des 
isom\'etries locales $\gg$ (\S\ref{s-isomet})
ont \'et\'e cr\'e\'ees pour de tels catalogues. 
Si les effets de l'\'evolution 
(et de l'angle de vue d'un objet comme un quasar) ne sont pas trop
forts, la premi\`ere technique est utilisable. 
La seconde m\'ethode est n\'ecessaire quand ces effets r\'eduisent 
significativement le nombre de paires d'images topologiques qui 
seraient vues s'il n'y avait pas de tels effets.

N\'eanmoins, l'existence d'un objet $\ll$ unique $\gg$ 
(puisque plus brillants
que tous les autres de leur classe) \`a une grande distance peut toujours
 \^etre utile dans les contraintes empiriques {\em contre} une topologie
non-triviale. Dans \S\ref{s-amasX}, l'usage des amas de galaxies 
s\'electionn\'es en rayons X est pr\'esent\'e. Dans ce contexte, 
un $\ll$ amas $\gg$ 
de galaxies doit \^etre consid\'er\'e plut\^ot comme un objet ponctuel que
comme un ensemble d'objets.

Finalement, la m\'ethode la plus \'el\'egante, la plus contraignante et
la plus s\^ure pour utiliser les informations bidimensionnelles du fond 
diffus cosmologique, c'est-\`a-dire qui proviennent (presque) de
la sph\`ere de l'horizon, est celle $\ll$ des cercles $\gg$ 
(\S\ref{s-cercles}).
Celle-ci va profiter des observations des satellites MAP et Planck
dont le lancement est pr\'evu  
pour l'an 2000 et l'an 2005 respectivement.

Par souci de simplicit\'e, le cas o\`u l'Univers est plat sans 
constante cosmologique ($\Omega_0=1$) est utilis\'e comme 
cas type ci-apr\`es. En r\'ealit\'e, une courbure hyperbolique est 
plus int\'eressante th\'eoriquement et observationnellement.

Si l'Univers est hyperbolique, les conditions initiales de l'inflation
doivent \^etre telles que l'Univers aujourd'hui soit presque plat (sur 
une \'echelle logarithmique), mais pas tout-\`a-fait. Si 
l'Univers \'etait multi-connexe avec une courbure n\'egative,
les g\'eod\'esiques seraient chaotiques et 
l'Univers primordial aurait eu ses g\'eod\'esiques $\ll$ mix\'ees $\gg$ telles 
que les conditions initiales pour l'inflation fussent remplies 
naturellement, \'evitant le probl\`eme de $\ll$ fine-tuning $\gg$ 
(\cite{Corn96a}~1996a).

En plus, si l'Univers est hyperbolique, $r_<$ et $r_>$ seront, 
 \`a quelques ordres de magnitude pr\`es, de l'ordre de $R_H,$ tandis
que $r_<$ et $r_>$ ne sont pas du tout contraints g\'eom\'etriquement
par rapport \`a $R_H$ si l'Univers est plat. 

Pour des discussions du cas hyperbolique, voir les travaux
pr\'esent\'es par \cite{Fag85}~(1985; 1989; 1996). Pour des 
repr\'esentations num\'eriques des vari\'et\'es de 
topologie hyperbolique non-triviale et des logiciels 
de visualisation, {\sc SnapPea} et {\sc geomview} \`a
http://www.geom.umn.edu/ sont conseill\'es.

\subsection{La cristallographie cosmique} \label{s-cristal}

Pour un catalogue d'objets qui ne souffrent pas trop d'effets d'\'evolution
et de l'invisibilit\'e d'\'eventuelles images topologiques, un histogramme 
des s\'eparations de paires d'objets pourrait montrer des pics dus
aux paires d'images topologiques s\'epar\'ees par des multiples 
des vecteurs qui $\ll$ g\'en\`erent $\gg$ la topologie (qui d\'efinissent 
le poly\`edre par rapport au rev\^etement universel). 
\cite{LLL96}~(1996) ont montr\'e par des simulations 
que cette m\'ethode devrait \^etre tr\`es 
efficace et ind\'ependante de la topologie, au moins si la courbure
de l'Univers est nulle. 

L'application de cette m\'ethode aux catalogues traditionnels des amas de
galaxies Abell et ACO, vus jusqu'aux d\'ecalages vers le rouge 
d'environ $z\approx 0\.25 \Rightarrow r\sim R_H/10,$ ne donne pas
de signal topologique. 

Ces catalogues ont \'et\'e s\'electionn\'es de 
fa\c{c}on subjective dans le visible. 
Il y aura bient\^ot des catalogues disponibles plus complets et homog\`enes 
\'etablis \`a partir des observations des satellites de rayons X, 
comme ROSAT, mais seulement \`a $z \ltapprox 1/2$ dans le futur proche. 
Un amas de galaxies est vu dans les rayons X par son
gaz chaud (\`a plus de $10^7$~K) et n'est pas soumis aux effets 
de projection qui compliquent la s\'election optique. Nous pourrions
esp\'erer donc que cette m\'ethode va augmenter les limites sur $2r_>$ 
(ou d\'etecter la topologie) aux \'echelles de plusieurs fois 
$R_H/10$ dans les prochaines quelques ann\'ees. 

\subsection{Recherche des isom\'etries locales} \label{s-isomet}

La limitation de la m\'ethode cristallographique aux catalogues complets
et aux objets peu assujettis aux effets de l'\'evolution emp\^eche son 
application aux objets vus jusqu'au plus que $R_H/2$ : les quasars.
Quoique l'\'evolution des quasars n'est pas encore bien comprise, 
les quasars doivent avoir des temps de vie tr\`es courts ou des
p\'eriodes d'activit\'e (donc de visibilit\'e) r\'ecurrentes. En plus, vu
d'un angle trop diff\'erent que celle de sa premi\`ere image topologique, 
un quasar appara\^{\i}tra beaucoup plus faible en luminosit\'e, 
par exemple comme une 
galaxie Seyfert, et aura une tr\`es petite probabilit\'e ($P \sim 10^{-8}$) 
d'\^etre d\'ej\`a observ\'e dans un catalogue.

Une approche est de consid\'erer des cas sp\'eciaux. \cite{Fag87}~(1987),
par exemple, ont cherch\'e des images de notre galaxie en tant que 
quasars s\'epar\'es par 180\deg ou 90\deg.

Une fa\c{c}on plus g\'en\'erale d'utiliser les quasars est de chercher 
les cas rares o\`u {\em plusieurs} objets dans deux images 
de {\em r\'egions \`a trois dimensions} sont \`a la fois visibles. 
C'est-\`a-dire, l'on cherche des isom\'etries des petites r\'egions
(de quelques centaines de \hMpc) dispers\'ees dans le rev\^etement 
universel. Une fois assez d'isom\'etries trouv\'ees, une topologie 
candidate peut \^etre test\'ee par sa pr\'ediction des positions 
(toujours \`a trois dimensions) d'images topologiques
d'autres quasars, ou par ses effets sur le fond diffus cosmologique,
par exemple.

\mbox{\protect\cite{Rouk96}~(1996)} 
a pr\'esent\'e cette m\'ethode et montr\'e son 
application au catalogue des $N\approx $~5000 
quasars avec $z>1$ d\'ej\`a observ\'es. Deux isom\'etries, c'est-\`a-dire 
deux paires de quintuples de quasars s\'epar\'es par plus de 
300\hMpc, sont trouv\'ees. Des simulations num\'eriques montrent qu'il y a
une chance de 30\% environ de trouver deux coincidences ou plus
selon les m\^emes crit\`eres si l'Univers est de topologie triviale.

Ces paires peuvent donc \^etre consid\'er\'ees au mieux comme 
d\'efinissant une candidate faible \`a la vari\'et\'e de l'Univers. Mais 
puisque les isom\'etries ainsi d\'efinies dans le rev\^etement universel
impliquent des r\'eflections, ce r\'esultat doit relancer le d\'ebat sur
la possibilit\'e que l'Univers est non-orientable.

Selon \cite{Chardin98}~(1998), un changement de l'orientabilit\'e 
d'une particule la transformerait de mati\`ere en anti-mati\`ere 
(ou l'inverse). Est-ce que les limites observationnelles sur
l'existence de l'anti-mati\`ere contraignent 
l'orientabilit\'e de l'Univers?

Puisque les quasars sont visibles jusqu'\`a $r \gtapprox R_H/2,$ 
les sondages futurs pourraient r\'ev\'eler la topologie de l'Univers
par cette m\'ethode d' $\ll$ isom\'etries locales $\gg$. Egalement, une
optimisation math\'ematique de cette technique ou l'application
de la version de \cite{Rouk96}~(1996) 
aux donn\'ees actuelles pour des courbures n\'egatives sont
des projets faisables sans attendre de nouveaux sondages.

\subsection{Amas en rayons X comme bougies standards} \label{s-amasX}

Pour qu'une des deux m\'ethodes pr\'ec\'edentes soit utilis\'es, 
il faut qu'un catalogue d'objets contienne non seulement leurs
positions (angles) c\'elestes, mais aussi leurs distances. 
Les distances sont bien estim\'ees (pour des objets \`a 
$r \gtapprox R_H/30$) au moyen de leurs d\'ecalages spectraux 
vers le rouge, 
$z.$ Mais il faut beaucoup plus de photons pour estimer $z$ que pour 
 \'etablir l'existence d'un astre sur une position angulaire. 

Avant qu'un catalogue dit $\ll$ photom\'etrique $\gg$ soit suivi par de la 
spectroscopie, l'utilisation de telles donn\'ees n\'ecessite le recours
 \`a la recherche d'images topologiques individuelles. 

Dans le cas des amas de galaxies vus par leur gaz chaud en rayons X,
la combinaison des d\'ecalages vers le rouge de quelques amas brillants
et lointains avec des limites photom\'etriques sur une grande partie
du ciel rendent possible une telle technique.

Puisque les amas massifs sont des objets qui viennent de s'effondrer
gravitationnellement tr\`es r\'ecemment, le temps d'\'echelle de leur
 \'evolution est comparable \`a l'\^age de l'Univers. Si on sais 
qu'un amas riche existe \`a un $z$ quelconque, il est tr\`es difficile de
voir comment les propri\'et\'es de cet amas peuvent changer de fa\c{c}on
significative aux \'epoques plus r\'ecentes, 
sauf si l'amas devient plus massif et plus lumineux.

Donc, \cite{RE97}~(1997) ont remarqu\'e que 
si l'amas le plus brillant que nous voyons est 
tr\`es \'eloign\'e 
de nous et s'il n'y a pas d'amas plus proches de luminosit\'es 
comparables \`a celui-ci, il ne peut pas y avoir d'images topologiques 
de cet amas distant qui sont plus proches \`a l'observateur que l'amas 
lui-m\^eme. Cette distance nous donnera donc une limite inf\'erieure \`a
 $r_<$ si les limites sur les amas les plus proches sont obtenues 
d'un sondage sur tout le ciel, 
ou dans le cas le plus vraisemblable, 
une limite inf\'erieure \`a $2r_>.$

Pour mettre des limites sur des images des amas \`a la fois
intrins\`equement plus brillants qu'une certaine 
luminosit\'e et plus proches d'une distance donn\'ee, les d\'ecalages 
spectraux sont inutiles.

Les amas les plus brillants connus, en combinaison avec des limites
 $\ll$ photom\'etriques $\gg$ sugg\`erent une limite de 
$2r_> \gtapprox $1000\hMpc$ \approx R_H/6$ (\cite{RE97}).

N\'eanmoins, une candidate faible pour la topologie de l'Univers a \'et\'e
trouv\'ee de mani\`ere s\'erendipiteuse parmi les amas les plus brillants : 
trois des sept amas \'etudi\'es forment un angle droit (\`a 2\% pr\`es) 
avec deux bras \'egaux \`a 1\% pr\`es. La topologie serait donc 
$T^2 \ttimes X$ o\`u $X$ est ind\'etermin\'e. \cite{RE97}~(1997) 
ont trouv\'e
des arguments contre l'identit\'e topologique de ces trois amas, 
mais la r\'efutation la plus concr\^ete serait la v\'erification de la 
non-existence des contre-images. 

Sp\'ecifiquement, si cette candidate \'etait la bonne, l'objet vu par ROSAT,
RX~J203150.4-403656, devrait \^etre un amas avec $0\.38 < z < 0\.40,$
et l'amas  AM~0750-490 (qui serait l'image topologique,
avec une valeur de $z$ faible, de 
 MS~1054-0321) serait \`a $0\.23 < z < 0\.26.$

\subsection{Fond diffus cosmologique : la m\'ethode des cercles} 
\label{s-cercles}

Il pourrait sembler moins efficace d'utiliser l'information 
distribu\'ee uniquement (presque) sur une sph\`ere \`a deux dimensions, 
mais le fait que le rayon du fond diffus cosmologique est presque
 \'egale \`a $R_H$ 
permet de surmonter cet obstacle. Plusieurs auteurs ont essay\'e 
d'utiliser les observations du fond diffus par le satellite COBE 
pour contraindre la topologie de l'Univers, mais avec des 
suppositions limitant la g\'en\'eralit\'e de leurs r\'esultats 
(voir \cite{Corn98}~1998 pour une revue).

Une m\'ethode qui sera valable pour une topologique quelconque 
(comme les m\'ethodes tridimensionnelles au-dessus cit\'ees), est celle
 $\ll$ des cercles identifi\'es $\gg$ de \cite{Corn96a}~(1996a). 

Pour comprendre cette m\'ethode, il faut \'etendre notre chiffonnement 
du rev\^etement universel par les copies du poly\`edre fondamental 
jusqu'\`a ce qu'on ait plusieurs copies du poly\`edre \`a l'ext\'erieur 
de la sph\`ere observable. Puisque chaque copie du poly\`edre est
 \'equivalente, on peut imaginer \`a la fois un observateur au centre
de la sph\`ere observable, et un observateur dans la m\^eme position 
physique dans une copie du poly\`edre \`a l'ext\'erieur de la sph\`ere 
(mais \`a moins de $2R_H$ du premier observateur).

Rappelons que l'univers physique consiste uniquement en le poly\`edre; 
le rev\^etement universel n'est qu'un espace apparent, mais utile pour
le calcul des g\'eod\'esiques. 

Maintenant, il faut ajouter la sph\`ere de l'univers observable 
du point de
vue du second observateur. 
L'intersection de cette sph\`ere avec la premi\`ere 
est un cercle. 
Les deux observateurs voient donc le long 
du cercle une suite de points d'espace-temps identiques : ce sont
les m\^emes points spatiaux et il faut l'\^age de l'univers pour que 
les photons arrivent \`a l'un ou \`a l'autre observateur. A condition
que le rayonnement sur ce fond diffus soit isotrope, autour du
cercle les deux observateurs voient la m\^eme suite de valeurs de 
l'exc\`es de temp\'erature au-dessus du corps noir ($\Delta T/T$). 

Mais les deux observateurs existent dans un seul univers physique. 
Ils sont \'equivalents \`a un seul observateur qui regarde le ciel dans 
(en g\'en\'eral) deux directions diff\'erentes mais en voyant les
m\^emes \'ev\'enements dans l'espace-temps.

Ind\'ependamment de la topologie, l'on verrait forc\'ement des cercles
identifi\'es sur le fond diffus si $r_> < R_H.$ Sauf si, par hasard, 
la Galaxie a une inclinaison particuli\`ere 
par rapport au 
poly\`edre fondamental, nous aurons \'egalement une limite sur
$r_<$ quand les observations du fond diffus cosmologique 
sont disponibles des satellites MAP et Planck.

\section{Conclusion}

Plusieurs m\'ethodes ont \'et\'e d\'evelopp\'ees pendant 
les cinq derni\`eres ann\'ees
pour d\'etecter ou contraindre la topologie de l'Univers. 
Elles pourraient \^etre appliqu\'ees aux 
nouveaux sondages d'objets astrophysiques et 
du fond diffus cosmologique dans la d\'ecennie \`a venir. 
En l'an 2008 au plus tard, 
nous devrions savoir si la topologie de l'Univers est 
observable ou non.

Les premi\`eres applications des m\'ethodes tridimensionnelles aux 
observations existantes ont d\'ej\`a fourni des param\`etres 
qui d\'eterminent des candidates 
faibles pour la vari\'et\'e (spatiale) dans laquelle nous vivons.
Ces candidates 
pourraient \^etre rapidement r\'efut\'ees (sauf si l'une ou l'autre
est la bonne) par des 
moyens observationnels modestes. De plus, les techniques pr\'esent\'ees 
ici n'ont pas encore \'et\'e utilis\'ees de fa\c{c}on 
compl\`ete sur les catalogues
de donn\'ees actuelles. Il reste donc plusieurs projets qui peuvent
\^etre r\'ealis\'es sans attendre les nouvelles donn\'ees observationnelles.


\end{document}